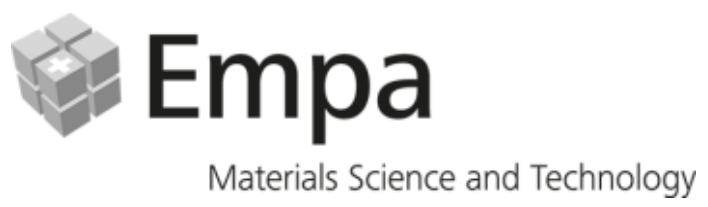

# Boron Co-Alloying in AlScN Wurtzite Ferroelectrics: Insights from an 850-Sample Combinatorial Study

*Federica Messi[1,2], Nathan Rodkey[1], Manuel Kober-Czerny[1], Sebastian Siol[1,*]*

[1] Laboratory for Surface Science and Coating Technologies, Empa − Swiss Federal Laboratories for Materials Science and Technology, Dübendorf, Switzerland
[2] Department of Materials, ETH Zürich, Zurich, Switzerland

*Corresponding author: sebastian.siol@empa.ch

Funding: SNSF, Swiss National Science Foundation (10004403)

## Abstract

AlScN wurtzite ferroelectrics are promising candidates for energy-efficient non-volatile memory. However, AlScN suffers from a high coercive field and reduced cycling endurance, and the limited tunability of its properties constrains further optimization. Co-doping AlScN with boron offers the promise of independently tailoring the chemical and structural properties, making AlScBN an attractive quaternary system. This material has already been explored for a few selected compositions, however, no systematic study of the full AlScBN compositional space exists. A combinatorial approach consisting of gradient deposition with HiPIMS at low temperatures of 250°C and automatic analysis of film properties allowed us to analyze a total of 850 unique samples within the AlScBN phase space. In addition to a full screening of the materials' chemical and structural properties, we fabricate and characterize combinatorial device libraries. XPS charge transfer analysis experimentally confirms that bond ionicity correlates with a reduction in the coercive field for AlScN and AlScBN systems, opposite trends are instead observed for AlBN. While the films maintain a high remanent polarization of 130–150 μC/cm$^2$, Sc and B co-doping reduces the coercive field from 7 MV/cm to 3 MV/cm. Notably, B co-alloying lowers the amount of Sc needed to lower the coercive field, reducing reliance on this scarce element. In addition, we find that co-alloying with B, notably improves cycling endurance, which is related to a reduction in defect density. These results establish AlScBN as a scalable, CMOS-compatible ferroelectric, positioning it as an interesting alternative to AlScN.

# 1. Introduction

Data generation is growing at an unprecedented rate, placing extraordinary demands on memory technology. Global data storage and processing already account for a significant and growing fraction of worldwide energy consumption, making the development of non-volatile, energy-efficient memory devices one of the central challenges of modern materials science. [1], [2] Ferroelectric materials, owing to their ability to store data without a constant power supply, are among the most promising candidates to address this challenge. Ferroelectric random-access memory (FeRAM) or ferroelectric field-effect transistors (FeFETs) promise fast switching, low power consumption, and high retention. [3], [4], [5] However, the requirement of low coercive fields (<1 MV/cm), high remanent polarization (>20 μC/cm$^2$), and high cycling endurance (>$10^7$ cycles) remains an open and critical challenge in the development of ferroelectric memory.

Aluminum nitride (AlN) based wurtzite ferroelectrics have recently emerged as promising candidates for ferroelectric non-volatile memory, due to their CMOS compatibility, high polarization and excellent retention.[1], [3], [4], [6] However, it was not until 2019, that ferroelectric switching was observed in AlScN. [7] Ferroelectric switching is facilitated by Sc substitution through coupled structural and chemical mechanisms. First, the large ionic radius mismatch between Sc (~0.885 Å) and Al (~0.535 Å) destabilizes the wurtzite structure toward a non-polar layered hexagonal transition polymorph, fundamentally lowering the c/a ratio and the internal u-parameter, inducing decrease in the switching energy barrier.[7], [8] Second, the lower electronegativity of Sc (1.36 on the Pauling scale, versus 1.61 for Al) enhances the bond ionicity of the metal-nitrogen bonds. This substitution replaces highly directional covalent $sp^3$ bonds with more compliant, non-directional ionic interactions, effectively lowering the elastic barrier to atomic displacement during switching. [9] While the role of bond ionicity has been explored through theoretical frameworks, direct experimental validation of how local electronic charge transfer correlates with macroscopic ferroelectric switching in these complex alloy systems remains elusive.[10] Together these effects reduce the coercive field with increasing Sc content. [7], [11], [12],[13] Separately, boron alloying of AlN in AlBN has been shown to moderately suppress the coercive field from ~6–7 MV/cm in AlN to ~5.2 MV/cm at B concentrations up to ~15 cat.% , while maintaining remanent polarization values of 125–136 μC/cm² and offering the additional benefit of a wider bandgap (>4.9 eV) that reduces dielectric losses and increases breakdown robustness [11], [12], [13], [14]. Despite these promising results, the mechanism by which boron facilitates ferroelectric switching is not fully understood as the underlying mechanisms differ from AlScN.

One main challenge in the development of AlScN is the limited thermodynamic solubility of Sc in AlN. The large ionic radius of Sc leads to high mixing enthalpies and and structural frustration for higher Sc content.[17] B, by contrast, has a significantly smaller ionic radius (~0.27 Å) than aluminum.[9] Co-alloying AlN with Sc and B is therefore expected to reduce compressive stress and facilitate higher non-equilibrium solubility of Sc, potentially enabling lower coercive fields while preserving high remanent polarization. Despite this promise, the AlScBN compositional space has received only limited attention to date. Yousefian *et al.* showed that B significantly suppresses leakage current, making this material system promising for high-temperature and high-field applications. [18] Subsequently, a combination of 10 cat. % B and 45 cat. % Sc has been reported to reduce the coercive field to as low as 2.5 MV/cm, though still with considerable leakage.[19] Finally, epitaxial AlScBN films have been grown on n-GaN, tuning the lattice mismatch by structural engineering, showing remanent polarization values in the range of 135–185 μC/cm² and coercive fields between 6.1 and 6.7 MV/cm for B/Sc compositions of 4/18, 6/18, and 6/9 cat.%. [20]

Existing studies cover only a narrow compositional window, leaving the interplay between B and Sc co-doping across the full relevant composition space essentially uncharted. A key obstacle to progress in the development of AlScBN is its compositional complexity. Individual sample-by-sample studies cannot capture the complex, coupled dependence of ferroelectric performance on both Sc and B content simultaneously. To address this challenge, we employ a combinatorial thin-film deposition approach, synthesizing a total of 850 unique AlScBN samples spanning a wide compositional space. Films are deposited by High Power Impulse Magnetron Sputtering (HiPIMS), a scalable, CMOS-compatible technique that affords precise control over ionic kinetic energy and enables highly compact, low-defect films even at low substrate temperatures. [6], [21], [22], [23] We have shown that HiPIMS enables consistent c-axis texture in wurtzite nitrides even during oblique angle deposition on a static substrate, which enables an evaluation of compositional effects without detrimental texture gradients during combinatorial thin film development. Our systematic structural characterization reveals that B incorporation preserves the pronounced c-axis texture essential for wurtzite ferroelectrics, while slightly modifying the lattice parameters and bond character of AlScN. Sc alloying leads to a change in c/a ratio whereas B substitution does not deform the lattice in the same way. Crucially, we utilize high-throughput X-ray photoelectron spectroscopy (XPS) to monitor the relative shift between the Al 2s and N 1s binding energies. This core-level shift serves as a direct proxy for bond ionicity, providing new insights into its evolution due to scandium and boron doping. Whereas the low electronegativity of scandium drives an increase in bond ionicity, boron doping suppresses it. At the device level, we demonstrate that B does not compromise the remanent polarization of AlScN, additionally reducing the coercive field at high Sc and B concentrations. These combined insights highlight that Sc and B lower the switching barrier in fundamentally different ways. Finally, B incorporation yields a noticeable improvement in cycling endurance, extending device lifetime to $10^7$ cycles without sample recharging, a benchmark that positions AlScBN as a promising candidate among wurtzite ferroelectrics for high-endurance non-volatile memory.

## 2. Results and Discussion

### 2.1. High-Throughput Compositional Exploration of Quaternary AlScBN

**Figure 1** presents the compositional space mapped across 850 samples from all deposited libraries, color-coded by library. The range investigated intentionally spans a broad region of the quaternary AlScBN system, extending beyond the known binary AlScN solubility limit. This wider exploration is motivated by the expectation that boron, due to its small atomic radius, relieves the lattice stress induced by Sc incorporation, potentially expanding the solubility limit in the quaternary system. Beyond this limit, Sc and B precipitate as rocksalt ScN and hexagonal BN respectively, degrading structural quality, increasing leakage current, and suppressing ferroelectric performance. [23] Within the solubility region, the absence of secondary phase precipitation defines a critical design window for achieving functional ferroelectric behavior, making compositional control essential for high-performance AlScBN devices.

Identifying the solubility boundary in quaternary AlScBN is inherently more complex than in ternary systems, as Sc and B contribute independently to lattice parameter evolution, requiring a two-dimensional compositional analysis to disentangle their effects. From our analysis, we estimate the solubility region for AlScN to extend up to approximately 30 cat.% and 20 cat.% in AlBN. For AlScBN the total cation substitution can reach up to 35%. This is supported by three converging indicators: an increase in FWHM values (**Figure 2b**), the loss of ferroelectric switching (**Figure 4a,b**), and deviations from Vegard's law linearity of the unit cell volume as a function of alloying concentration (**Figures S1 and S2**). [24]

**Figure 1**. Ternary diagram of Al-Sc-B compositional space showing 850 combinatorial library samples (each library with a different marker color).

## 2.2. Structural Characterization

First, we turn to the crystallographic quality of the AlScBN films. In the XRD 2theta-omega measurements a prominent peak at 36° (002, **Figure 2a**) can be seen, which confirms the preferential c-axis orientation and the absence of differently oriented grains for all samples on the device libraries. (**Supplementary Figure S2**) Rocking curve measurements focused on the (002) peak are used to determine the degree of out-of-plane texture or misorientation within the film. Our films show FWHM values below 3.5° for the majority of the samples, consistent with state-of-the-art reports for AlScN polycrystalline thin films, with higher values observed towards the edges of the expected solubility window. [25], [26] (**Figure 2b**, **Figure S3**)

Interestingly, AlBN shows very low FWHM values, around 2°, even at high B composition compared to literature. [8]

XRD data were further utilized to extract the a and c lattice parameters and unit cell volume of the wurtzite phase. The results of the analysis are in good agreement with DFT trends previously reported for AlScN, AlScBN and AlBN. [8], [27] These parameters are fundamental indicators of the lattice distortion induced by the doping atoms, which directly correlate with ferroelectric performance, importantly a decrease in the c/a ratio induces a decrease in the switching energy barrier. As expected from literature, in AlScN and AlScBN, the c-parameter decreases with increasing Sc alloying, while the a-parameter increases, leading to a desired decrease in the c/a ratio, as shown in **Figure 2c** and **Figure S4**. In AlBN, by contrast, both c- and a-parameters decrease with B content, resulting in an unchanged c/a ratio. While this would not inherently lower the energy barrier for polarization reversal, DFT calculations predict a decrease in the internal parameter u, which describes the relative displacement of the nitrogen and metal sublattices along the c-axis. This reduction is understood to be a primary driver of ferroelectric switching in this material system. [8] AlScBN c/a ratio values sit between the AlScN and AlBN datasets, allowing compositional tunability of the lattice properties. This result is also observable in the unit-cell volume trends shown in **Figure 2d**. In AlBN, the volume decreases slightly with B incorporation, consistent with the smaller ionic radius of B relative to aluminum. In AlScN, the large atomic radius of Sc drives a more substantial volume increase. Notably, the introduction of B in AlScN does not measurably alter this volume trend: the volume evolution of AlScBN follows that of AlScN.

**Figure 2**. Structural characterization of AlScBN, AlBN, and AlScN across the full combinatorial compositional space. **(a)** Representative 2theta-omega diffractograms confirming exclusive c-axis texture in AlScBN films across all compositions; the peak near 40° corresponds to the Pt bottom electrode. **(b)** Rocking curve FWHM map of the (002) reflection across the AlScBN compositional space, with the majority of samples showing values below 3.5°. **(c,d)** c/a ratio, and unit cell volume as a function of total doping content for all three material systems. The dashed vertical line marks the AlScBN est. solubility limit (~33 at.%) visible by an increase in the spread distribution of the data. AlScBN structural parameters sit between AlScN and AlBN trends, confirming compositional tunability. The absence of B-driven volume change in AlScBN raises questions regarding the structural role of B in the quaternary lattice.

## 2.3 Bond Ionicity Analysis

Bond ionicity is an important factor in determining the switching barrier in wurtzite ferroelectrics.[10], [28] Increasing bond ionicity lowers the ferroelectric switching energy barrier by reducing bond strength. [10] Combining the extensive compositional and structural characterization with high-throughput XPS mapping enables a systematic chemical state analysis of the constituent atoms. Since AlN-based thin films are highly insulating, measured absolute binding energies can be affected by many experimental factors such as differential charging or a lack of an appropriate charge reference.[29], [30], [31], [32] A practical and powerful alternative is charge transfer analysis in which only the energetic distance between two core level photoemission lines is evaluated.[32], [33], [34] In this case we are focusing on the relative shift between the Al 2s and N 1s binding energies, which serves as a proxy for bond ionicity, providing new insights into its evolution within mixed-cation systems.

In XPS, core-level binding energies are sensitive to the local chemical environment of an atom and the degree of electron transfer between bonded species. In an ionic bond, valence electron density is transferred from the cation toward the more electronegative anion. This reduced electron density leaves the cation's core electrons less shielded from the nuclear charge, both in the ground state and during photoemission, shifting its core-level binding energies to higher values. Conversely, the anion gains electron density, resulting in enhanced screening and a shift toward lower binding energies. In AlN and its alloys, the Al–N bond is partially ionic, and the binding energy difference between the N 1s and Al 2s levels ($\Delta BE_{A\text{-}N} = BE(N\ 1s) - BE(Al\ 2s)$) directly reflects this charge transfer: as ionicity increases, Al loses more electron density (Al 2s shifts to higher BE) while N gains more (N 1s shifts to lower BE), causing ΔBE to decrease. ΔBE therefore provides a semi-quantitative probe of bond ionicity that avoids the uncertainties associated with absolute binding energy calibration and differential charging, since only relative shifts are used.[[33], [35], [36]

The incorporation of Sc, which compared to Al has lower electronegativity increases the overall cation-anion bond ionicity, results in a measurable decrease in this binding energy difference, as confirmed in **Figure 3**. (See **Figure S9c**, for the full dataset showing bond ionicity versus coercive field.) This relationship between bond ionicity and Sc doping is apparent in both AlScBN and AlScN. Both systems exhibit a difference between the N 1s and Al 2s peaks decreasing from approximately 278.2 eV at 5% Sc down to 277.75 eV at ~37.5% Sc.

While in AlScN the reduction of the switching energy barrier is associated with both increased bond ionicity and lattice distortion, the mechanism in AlBN appears fundamentally different. B has higher electronegativity than Al, so its introduction increases bond covalency (**Figure**

**3b),** and the reduction of the coercive field is therefore not driven by bond ionicity. Instead, the reduction of the u-parameter facilitates displacement of B atoms from the nitrogen plane, promoting ferroelectric switching. Furthermore, B has a strong tendency towards three-fold coordination. Consequently, B in the wurtzite AlBN structure is predisposed to migrate towards a layered hexagonal transition phase, reducing the energy for ferroelectric switching and therefore the coercive field. This has been hypothesized previously by Hayden et al. and our results on bond covalency and c/a ratio variation, support this observation. [8] In AlScBN, bond ionicity is governed by Sc content, and no systematic dependence on B concentration is apparent (**Figure 3b**). Taken together, the reduction of the coercive field in AlScN and AlScBN is governed by three distinct mechanisms: structural distortions arising from changes in the c/a ratio and the u-parameter, and a chemical contribution from changes in bond ionicity.

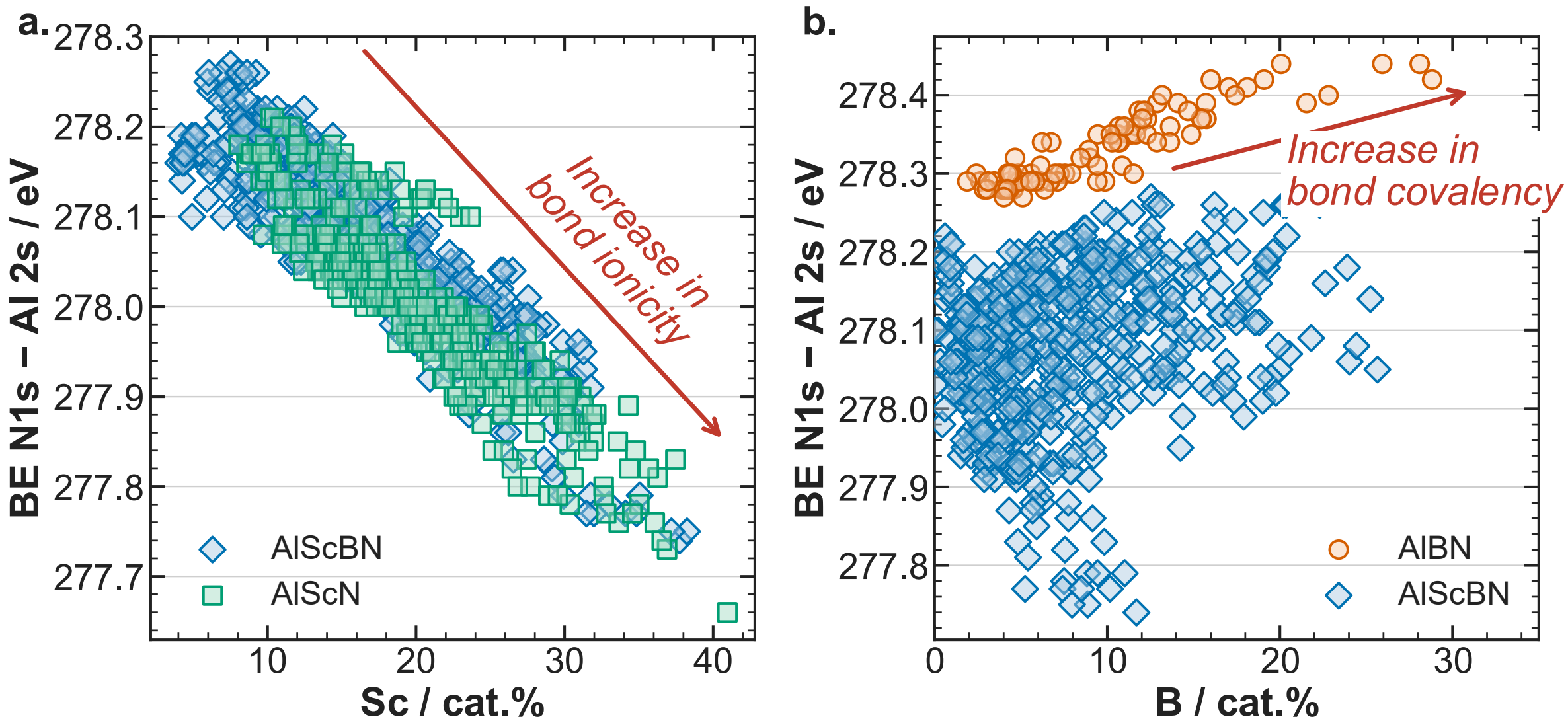

**Figure 3**. Bond ionicity analysis across the AlScBN compositional space. N 1s – Al 2s binding energy difference as a measure of bond ionicity plotted against **(a)** Sc content for AlScBN and AlScN, and **(b)** B content for AlBN and AlScBN. In panel **(a)**, bond ionicity increases systematically with Sc content in both AlScBN and AlScN, with both systems showing comparable trends, confirming Sc as the primary driver of bond ionicity modification in the quaternary system. In panel **(b)**, AlBN forms a distinct cluster at higher binding energy differences than AlScBN, reflecting increased bond covalency upon B incorporation and confirming a fundamentally different $E_c$ reduction mechanism in AlBN compared to AlScN. No systematic variation with B content is observed in AlScBN.

### 2.4. Device Characterization

The influence of B on the ferroelectric properties of AlScBN is of particular interest in assessing the degree to which co-doping can improve upon the baseline AlScN system.

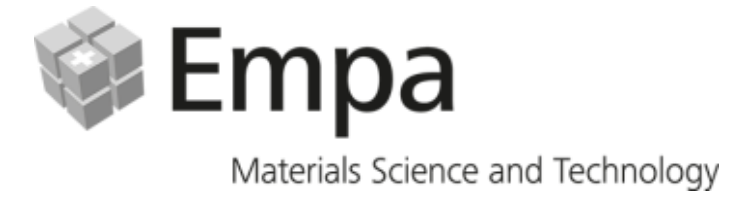

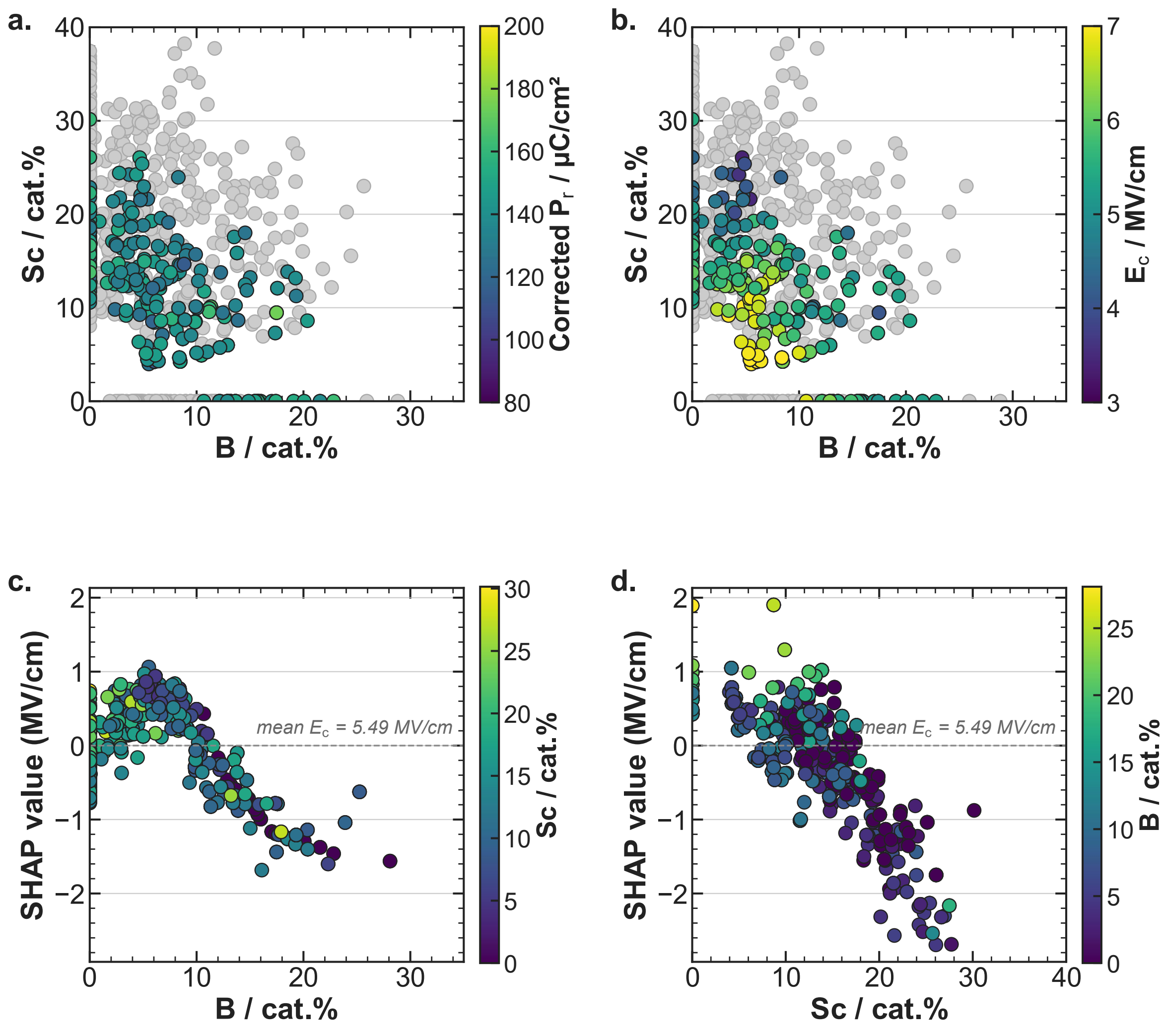

**Figure 4.** Ferroelectric performance maps and SHAP feature attribution for the coercive field $E_C$. **(a, b)** Composition maps (B vs Sc content) for the ferroelectric devices, coloured by **(a)** corrected remanent polarisation $P_r$ (130–150 µC/cm²) and **(b)** coercive field $E_C$ (3–7 MV/cm). Grey circles in the background indicate the samples for which no ferroelectric switching was observed. **(c, d)** SHAP (SHapley Additive exPlanations) dependence plots showing the contribution of each compositional feature to the predicted $E_C$, relative to the mean $E_C$ = 5.51 MV/cm (grey dashed line). In **(c)**, SHAP values are plotted against B content and colored by Sc content; in **(d)**, against Sc content and colored by B content. SHAP is a statistical model that deconvolutes inputs (B and Sc) from a target output ($E_c$). Positive SHAP values indicate alloy concentrations that increase $E_C$ above the mean of the dataset; negative values indicate a reduction. The SHAP analysis indicates that B initially increases the $E_c$, before leading to significant reductions. Notably, reductions in $E_c$ appear to be relatively linear and of similar magnitude between Sc and B

Automated PE loop screening was employed to characterize approximately 400 ferroelectric devices with varying composition, evaluating the remanent polarization $P_r$ and coercive field $E_c$. Representative PE loops for all three material systems are shown in **Figure S6**. All reported remanent polarization values are corrected for leakage current through PUND measurements, as shown in **Figure S7**. As depicted in **Figure 4a**, both AlScN and AlScBN thin films exhibit $P_r$ values consistently in the range of 130–150 µC/cm². This value appears to be largely independent of the composition. Overall these results are consistent with our previous work on HiPIMS of AlScN and suggest that $P_r$ degradation at high doping levels is governed by structural frustration rather than chemical changes. [13] The high adatom mobility inherent to the reactive HiPIMS process enables consistent c-axis texture over wider compositional ranges up to the expected solubility region, allowing for achievement of high $P_r$ even at high doping concentrations.[13], [23], [37] (**Figure S8**)

Regarding the coercive field, the co-incorporation of Sc and B facilitates a significant reduction in the coercive field, yielding values as low as ~3 MV/cm (**Figure 4b**). Coercive fields comparable to those of high-Sc AlScN (~25 cat.%) can be achieved at significantly lower Sc content when B is introduced (~10 cat.% Sc, ~13 cat.% B), suggesting B as a lever for reducing $E_c$ without excessive Sc loading.

Looking at the ternary AlBN in **Figure S8a**, our results confirm what has already been established in the literature, a saturation of the coercive field around 15% B doping. Higher B content induces severe structural frustration, triggering local phase segregation toward the h-BN phase and mitigating the benefits originally driven by boron alloying. (**Figure S8**). [8]

Subsequent we used Shapley additive explanations (SHAP) as a statistical tool to visualize component trends and to deconvolute the effect of inputs (Sc and B) to a model output ($E_c$). More information about this analysis is given in the methods section. In **Figure 4c** and **d** the SHAP values for B and Sc concentrations can be found respectively. The dashed line at 0 represents the mean of the dataset, and positive SHAP values indicate compositions where the B or Sc content are estimated to increase $E_c$ over the mean, and negative to decrease. For B, an initial increase in $E_c$ is observed followed by a significant decrease and possible saturation at ~20% B. This can be visualized as well in **Figure 4b**. Notably, both Sc and B decrease $E_c$ linearly and by similar magnitudes.

## 2.5. Cycling endurance and defects

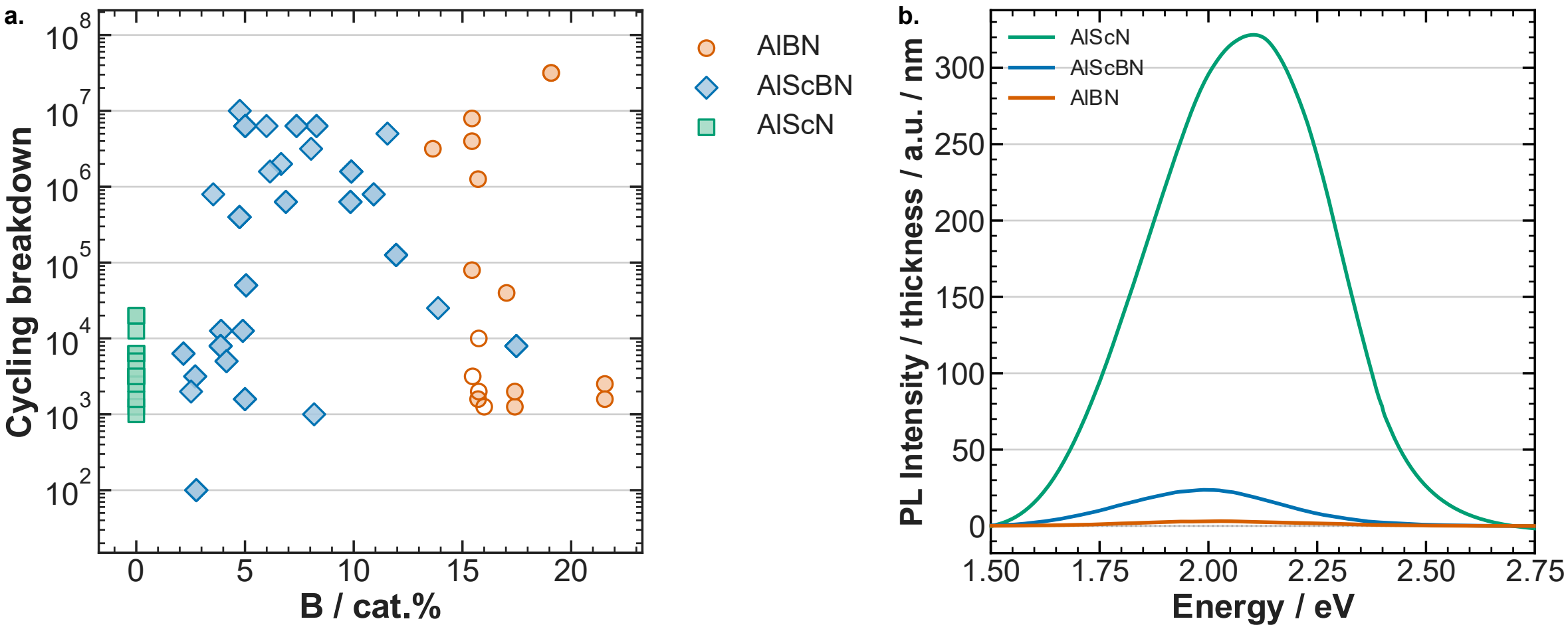

**Figure 5**. **Relationship between cycling endurance and defects. (a)** Cycling breakdown as a function of boron content for AlBN, AlScBN and AlScN devices. Data is pre-filtered by XRD FWHM < 3° and thickness range 220 – 270 nm. AlScN points are plotted at B = 0 cat.%. Filled markers indicate samples that failed via dielectric breakdown, while empty markers indicate samples that degraded due to fatigue-induced loss of remanent polarization, which was observed only for AlBN samples. Boron introduction induces an improvement in cycling endurance from $10^4$ to $10^7$ cycles. **(b)** PL spectra of selected AlScN (23 cat. % Sc), AlScBN (24 cat. % Sc, 8 cat. % B), and AlBN (14 cat. % B) normalized over the thickness of the sample. The peak intensity is maximum for AlScN samples and minimal for AlBN samples. Interestingly AlScBN samples show intermediate intensity at the same scandium load as AlScN. The decrease in the PL area proves the decrease of defect density due to boron doping.

Cycling endurance was evaluated to assess the operational lifetime of the ferroelectric devices. A distinct divergence in the dominant failure mechanisms was observed between the material systems: AlScN and AlScBN devices typically undergo dielectric breakdown via short-circuiting, consistent with thermal run-away caused by nitrogen vacancy-driven degradation. [38], [39] In contrast, AlBN devices exhibit both failure modes, short-circuiting and polarization fatigue, with fatigue being characterized by a gradual decline in remanent polarization rather than catastrophic failure. (**Figure S9**) Notably, B incorporation significantly enhances endurance, with AlScBN and AlBN films sustaining up to $10^7$ cycles without breakdown (**Figure 5a**, **Figure S10**). These results match the highest values reported in literature for fully switching cycling experiments in single-layer AlN-based thin films. [40] Crucially, these benchmarks were achieved via combinatorial deposition without optimizing process parameters for specific compositions, highlighting the clear potential to achieve even higher cycling endurance through targeted process optimization.

Given that nitrogen vacancies are considered the primary drivers of dielectric breakdown in these nitrides, PL spectroscopy is performed to evaluate the defect density in the cycled devices. **Figure 5b** shows the PL spectra of representative cycled AlScN, AlScBN and AlBN, normalized by the thickness to mitigate its influence on the data. The PL intensity is substantially reduced in AlBN and AlScBN relative to AlScN, a trend confirmed by the integrated peak area, which serves as a proxy for defect density. These results suggest that the improved endurance is associated with B co-doping mitigating the defects induced by Sc incorporation. [38], [41] However, further investigation is required to confirm this mechanism and identify the precise defect species being suppressed.[39], [42]

## 3. Conclusion

High-throughput combinatorial methodology, linking composition-gradient film depositions with automated characterization across 850 distinct samples, was successfully applied to map the quaternary AlScBN material space. We map the entire solubility region for single-phase wurtzite material and find that for AlScBN the total cation substitution can reach up to 35%. The solubility was estimated based on a map of the FWHM of the (002) rocking curve, which deteriorates abruptly in the presence of secondary phases. In addition, deviations from Vegard's law linearity of the unit cell volume can be observed for higher alloying concentrations. XRD analysis confirms that the deposited films exhibit both high crystalline quality and strong c-axis orientation up until the expected solubility window. Furthermore, the c/a ration of AlScBN scales predictably between those of the AlScN and AlBN, demonstrating compositional tunability of the lattice parameters.

XPS charge transfer analysis serves as a proxy for bond ionicity, with the relative binding energy shift between the Al 2s and N 1s core levels tracking changes in the local electronic environment. This study experimentally shows that increasing Sc content progressively increases the bond ionicity. In contrast, B substitution drives the bonding character towards increased covalency.

At the device level, the remanent polarization remains unaffected by varying doping ratios and total composition, while B incorporation enables lower coercive fields at equivalent Sc content. Two distinct mechanisms contribute to coercive field reduction in AlScBN. Both Sc and B alloying reduce the potential energy barrier for polarization reversal. For B, the reduction is driven by structural phase competition between the wurtzite host and the equilibrium layered hexagonal BN, which brings the non-polar hexagonal intermediate closer in energy to the two polar wurtzite states. For Sc, the lattice distortion induced by the large atomic radium of the dopant induces a decrease in the c/a ratio, which distorts the crystal lattice, forcing it to approach the non-polar hexagonal transition plane. [43] Additionally, Sc-driven bond ionicity increase reduces the directionality of the Al–N bonds, providing a further independent contribution to barrier reduction. Most strikingly, B incorporation improves cycling endurance to $10^7$ cycles, likely through a reduction in nitrogen vacancy concentration. Short-circuit failure, consistent with nitrogen vacancy accumulation, remains the dominant breakdown mechanism, though its frequency appears reduced with BN incorporation. [38], [41] Further studies are required to address the type and the nature of defect suppression by boron integration in the AlScN lattice. These results demonstrate that AlScBN deposited by HiPIMS at 250°C is a CMOS-compatible, scalable material of genuine interest for non-volatile memory applications, combining exceptional crystal quality with improved cycling endurance. Transitioning from these combinatorial layers to targeted single-composition devices with optimized process parameters provides a clear trajectory to push endurance limits for wurtzite nitride ferroelectrics.

## 4. Experimental Methods

### 4.1. Sample Preparation

For the purpose of this study, a total of 19 combinatorial materials libraries, comprising 850 individual samples, were synthesized via combinatorial sputtering deposition. The libraries featured different elemental compositions, specifically 2 AlBN, 5 AlScN, and 12 AlScBN libraries. Depositions were performed in a UHV deposition chamber (AJA International, ATC-1800). P-type (001) 2 × 2 $in^2$ silicon wafers were used as substrates, and a 9 × 5 grid was utilized to define 45 discrete samples on each materials library.
Prior to deposition, the substrates were prepared with an RF etching step. This was followed by the deposition of a base stack consisting of an RF nitridation of the clean Si wafer, a 10 nm DC-sputtered Ti layer (serving as an adhesion layer and diffusion barrier), and a 200 nm Pt back contact.
For the main film depositions, two or three sputter guns were used depending on the target material system (AlBN, AlScN, or AlScBN). The substrates were kept stationary during the process to deliberately induce thickness and compositional gradients across the library. The elemental targets included Al (2 inch diameter, HMW Hauner, 99.9999 at.% purity), Sc (2-inch diameter, Plasmaterials, 99.9 at.% purity, oxygen content <1000 ppm), and BN (2 inch diameter, Kurt J. Lesker, 99.5 at. % purity). All layers were deposited at a temperature of 250 °C and a working pressure of 3 µbar, with gas flows set to 20 sccm for Ar and 12 sccm for $N_2$. Various deposition configurations were employed depending on the library. The AlBN libraries were deposited using High-Power Impulse Magnetron Sputtering (HiPIMS) on the Al target and RF power on the BN target. The AlScBN libraries were similarly deposited using HiPIMS on both Al and Sc, and RF power on BN. For the AlScN libraries, three distinct configurations were explored: both Al and Sc on HiPIMS; Al on RF and Sc on HiPIMS; or both Al and Sc on RF. When HiPIMS was utilized, the pulses on the active guns were synchronized. Power was applied to the targets in 10 µs pulses at a varying frequency to maintain a peak current of 5 A for Sc and 8 A for Aluminum. Additionally, two of the AlScBN libraries were deposited with an applied substrate HiPIMS bias of -10 V and -20 V, respectively. This applied bias featured a pulse length of 40 µs and was delayed by 18 µs with respect to the gun pulses, synchronizing it with the arrival of metal ions at the substrate. Accelerating just the metal-ions leads to improved ad-atom mobility while minimizing detrimental Ar-ion bombardment. [37], [44] This timing evaluation was conducted using gated mass spectroscopy (Hiden EQP 300) to study the time of flight of different plasma ion species, as detailed in our previous work.[45]

Finally, top contacts with diameters of 100, 200, 400 and 600 µm were deposited using either Pt or W in a sputter-down configuration via shadow masking. The pad area was later measured in an optical microscope and corrected to avoid overestimating the remanent polarization.

### 4.2. Sample Characterization

X-ray diffraction (XRD) analysis was performed using a Bruker D8 diffractometer in Bragg-Brentano geometry with Cu-Kα radiation. 2theta-omega scans covering a 2θ range of 30°–45° with a step size of 0.01° were used to assess crystal orientation via the (002) reflection. Lattice parameters a and c were extracted from a second 2theta-omega scan conducted with a ψ tilt of 42°, covering a 2θ range of 45°–55° to capture the (102) reflection. Rocking curve measurements of the (002) peak were carried out over an ω range of 12°–24° to quantify c-axis texture. Film thickness was determined using a Dektak profilometer. (See **Figure S3**)

Surface composition was determined by X-ray photoelectron spectroscopy (XPS) using a PHI Quantera instrument with monochromated Al Kα radiation. All spectra were referenced to the adventitious carbon C 1s peak at 284.8 eV leading to a typical inaccuracy of ± 0.2 eV.[30] Importantly in this paper we only report relative binding energy shifts, which are insensitive to these accuracies and differential charging.[32] Peak fitting was performed using a Lorentzian-Gaussian line shape with Shirley background subtraction for all elements except nitrogen, for which a Tougaard background was applied. Cation ratios were calculated from the integrated areas of the Al 2s, Sc 2s, and B 1s peaks after applying the respective relative sensitivity factors. Bond ionicity was evaluated from the binding energy difference between the N 1s and Al 2s peaks, which provides a consistent reference across all compositions as aluminum is present in all samples (see **Figure S5** for XPS peak fitting).

The ferroelectric properties were measured across all library samples using automatic mapping in a double beam laser interferometer (DBLI) equipped with an automatic probe station and a thin film analyzer(TF 2000, aixACCT Systems). Dynamic hysteresis measurements were performed by applying a voltage in the range of 120–200 V at a frequency of 3 kHz. The coercive field was defined as the average of the positive and negative fields at which the current density reaches its maximum, and the remanent polarization as the average of the positive and negative polarization values at zero applied field. Polarization values were corrected for leakage current contributions using positive-up-negative-down (PUND) measurements, which isolate the switching current from the leakage current (see **Figure S7**). Cycling endurance was assessed by fatiguing samples at 100 kHz at the coercive field and monitoring the evolution of polarization as a function of cycle number (full switching). Breakdown was defined as either dielectric short-circuit failure or a reduction of the remanent polarization below 20 $\mu C/cm^2$.

PL measurements were performed using a confocal Raman microscope (WITec alpha300 R) with 10x magnification. The laser excitation wavelength was 488 nm with an estimated spot size of 1 µm at the sample surface. The power at the laser was set to 20 mW, however, the exact power density at the sample is unknown. The beam path within the microscope was aligned using a Si reference sample. For each sample, first a z-alignment scan was done to find the position, where the total area of the PL was maximized. Then a PL spectrum was recorded with a 6s integration time and five averaged to yield a good signal-to-noise ratio. All measurements were performed in ambient conditions.

### 4.3. Data Processing

Shapley additive explanations (SHAP) was popularized by Lundberg et al. in 2017[46] and is described in greater detail in a previous work by our group.[47] For the analysis a Gaussian Proccess Regressor (GPR) based on a Matérn kernel was used as the surrogate model, and hyperparameters tuned were: the signal variance, two per-feature Matérn length scales (B and Sc), and a white-noise variance which were optimized by maximizing the log marginal likelihood with the L-BFGS-B optimizer, using 20 random restarts to mitigate local optima; the Matérn smoothness was fixed at $\nu = 5/2$.

## Acknowledgements

F.M. and N.R. acknowledge funding from the Swiss National Science Foundation (SNSF, Project No. 10004403). M.K.-C. acknowledges funding from the Swiss National Science Foundation (SNSF, Project No. 226588). The authors further acknowledge Morgan Trassin at ETH for helpful discussions as well as Stefanie Frick and Kerstin Thorwarth for their assistance with XPS analysis.

## Author contributions

Federica Messi: Conceptualization, Investigation, Formal analysis, Visualization, Writing – original draft. Nathan Rodkey: Investigation, Formal analysis, Writing – review & editing. Manuel Kober-Czerny: Investigation, Formal analysis, Writing – review & editing. Sebastian Siol: Conceptualization, Supervision, Methodology, Formal analysis, Funding acquisition, Writing – review & editing.

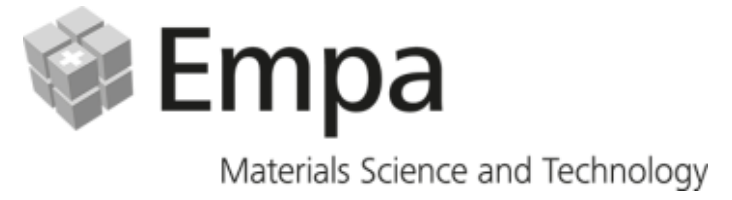

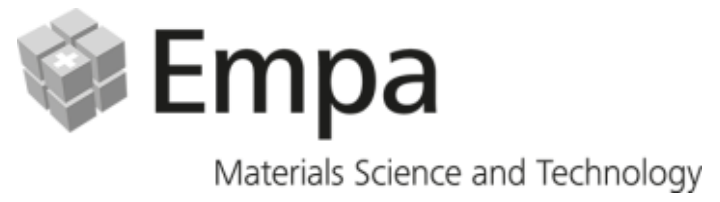

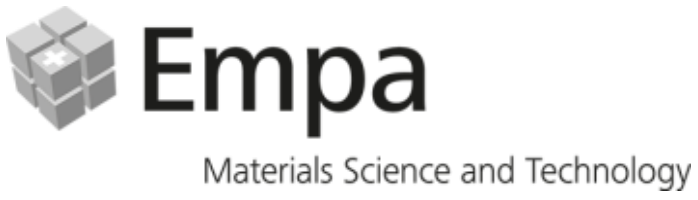

Supporting Information

# Boron Co-Alloying in AlScN Wurtzite Ferroelectrics: Insights from an 850-Sample Combinatorial Study

*Federica Messi[1,2], Nathan Rodkey[1], Manuel Kober-Czerny[1], Sebastian Siol[1, *]*

[1] Laboratory for Surface Science and Coating Technologies, Empa – Swiss Federal Laboratories for Materials Science and Technology, Dübendorf, Switzerland
[2] Department of Materials, ETH Zürich, Zurich, Switzerland
*sebastian.siol@empa.ch

Funding: SNSF, Swiss National Science Foundation (10004403)

Keywords: HiPIMS, ferroelectrics, AlScBN, memory, nitrides, AlScN, high throughput experiments, combinatorial screening, thin films

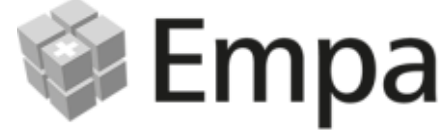

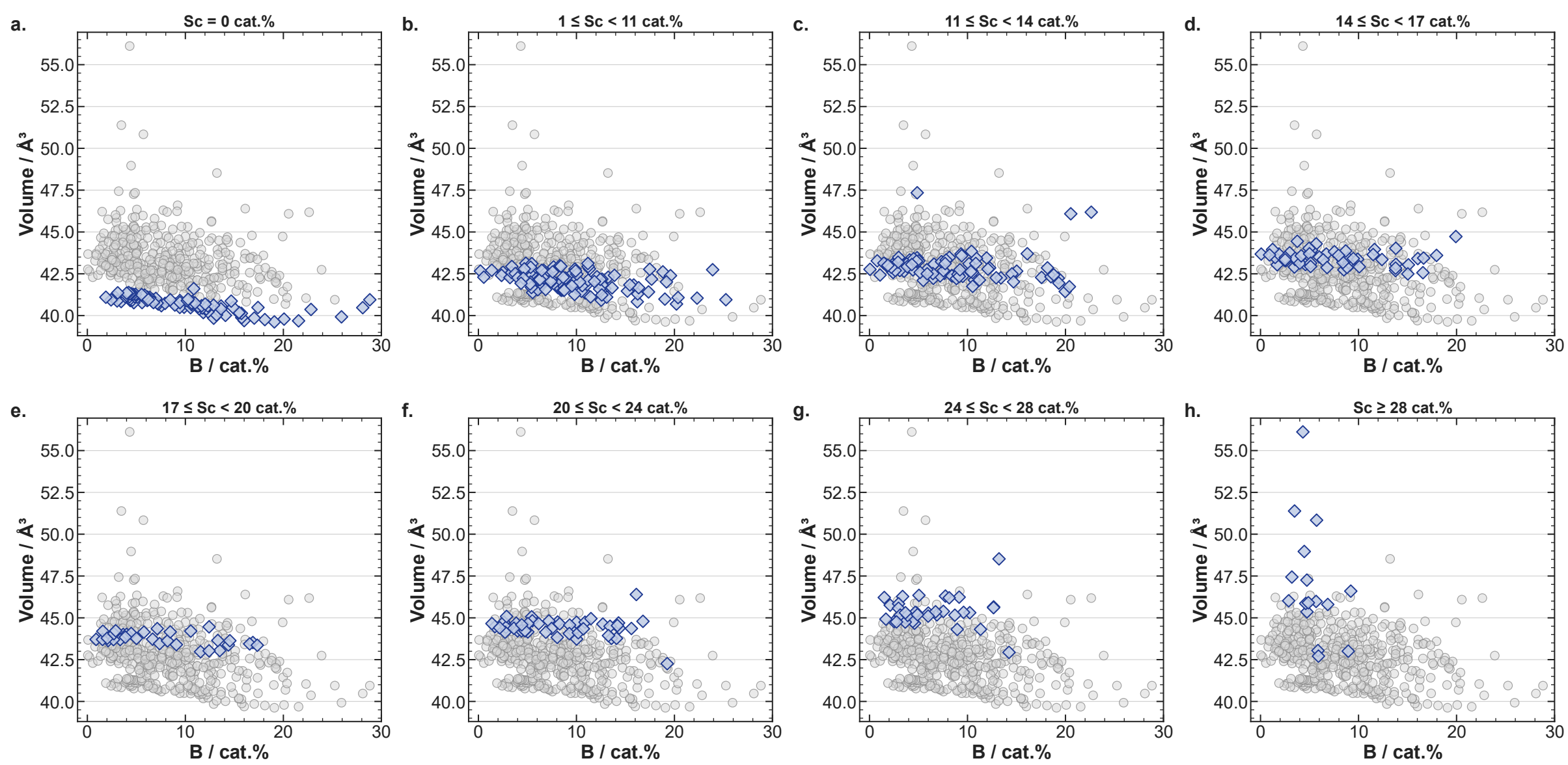

**Figure S1.1. Solubility limit evaluation from unit-cell volume.** Vegard's-law analysis of the unit-cell volume as a function of B content for AlScBN thin films, sliced by Sc concentration. Each panel displays the volume of all AlScBN samples with the subset falling within the indicated Sc range highlighted. Deviation of the volume from linear Vegard scaling signals the onset of wurtzite phase decomposition. The Sc concentration ranges are: (a) Sc = 0 cat.%, (b) 1–11 cat.%, (c) 11–14 cat.%, (d) 14–17 cat.%, (e) 17–20 cat.%, (f) 20–24 cat.%, (g) 24–28 cat.%, and (h) Sc ≥ 28 cat.%.

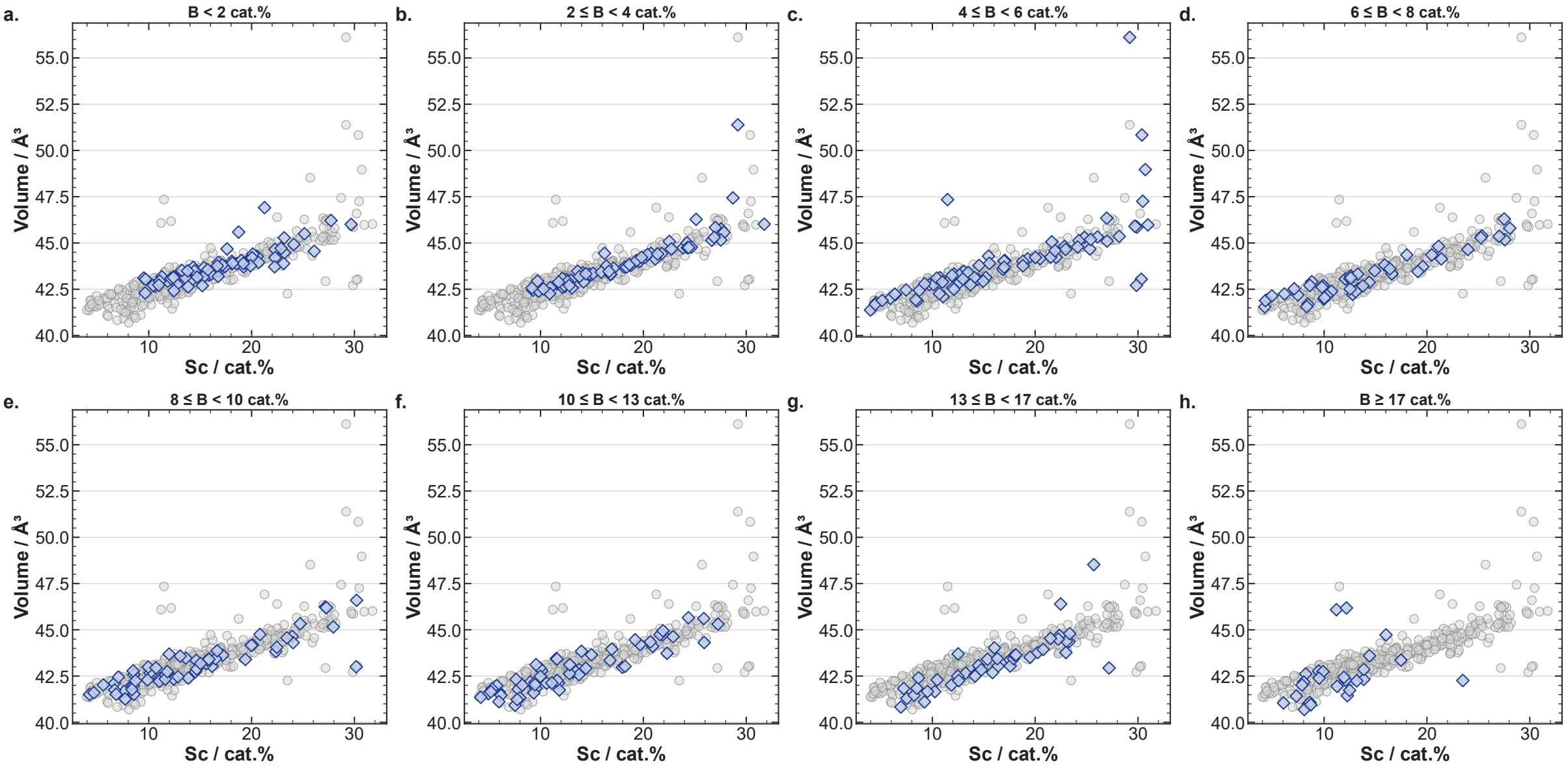

**Figure S1.2. Solubility limit evaluation from unit-cell volume.** Vegard's-law analysis of the unit-cell volume as a function of Sc content for AlScBN thin films, sliced by B concentration. Each panel displays the volume of all AlScBN samples with the subset falling within the indicated B range highlighted. Deviation of the volume from linear Vegard scaling signals the onset of wurtzite phase decomposition. The B concentration ranges are: (a) B < 2 cat.%, (b) 2–4 cat.%, (c) 4–6 cat.%, (d) 6–8 cat.%, (e) 8–10 cat.%, (f) 10–13 cat.%, (g) 13–17 cat.%, and (h) B ≥ 17 cat.%. In panel a the solubility limit is extrapolated from a previous study by Patidar et al. [37]

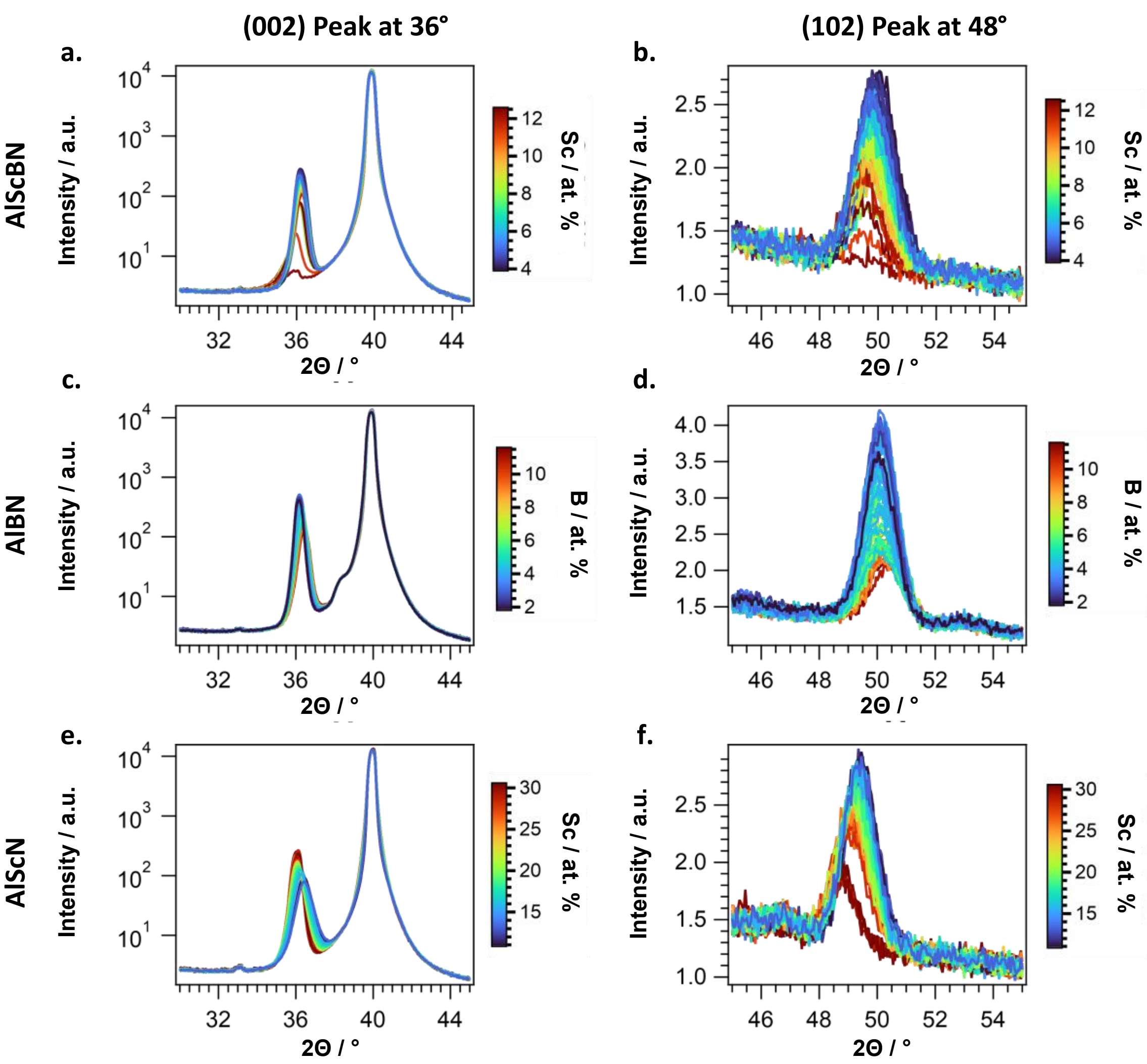

**Figure S2. XRD patterns for (002) and (102) peaks**. Representative XRD 2theta-omega diffractograms for all three material systems. Left column: symmetric 2theta-omega scans showing the wurtzite (002) reflection near 36° for AlScBN (a), AlBN (c), and AlScN (e), plotted on a logarithmic intensity scale. The peak near 40° is attributed to the Pt (111) reflection from the bottom electrode. The absence of additional film reflections confirms exclusive c-axis texture across all compositions. Right column: asymmetric 2theta-omega scans acquired with a ψ tilt of 42°, showing the (102) reflection near 48–50° for AlScBN (b), AlBN (d), and AlScN (f). The systematic shift of the (102) peak position with doping concentration enables extraction of both a and c lattice parameters. Individual scans are color-coded by Sc content (AlScBN, AlScN) or B content (AlBN) as indicated by the respective colorbars. Each panel overlays representative scans drawn from across the full compositional range of the corresponding library.

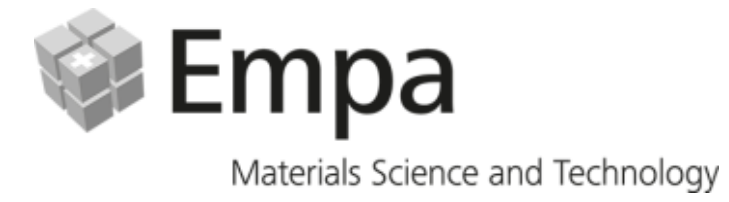

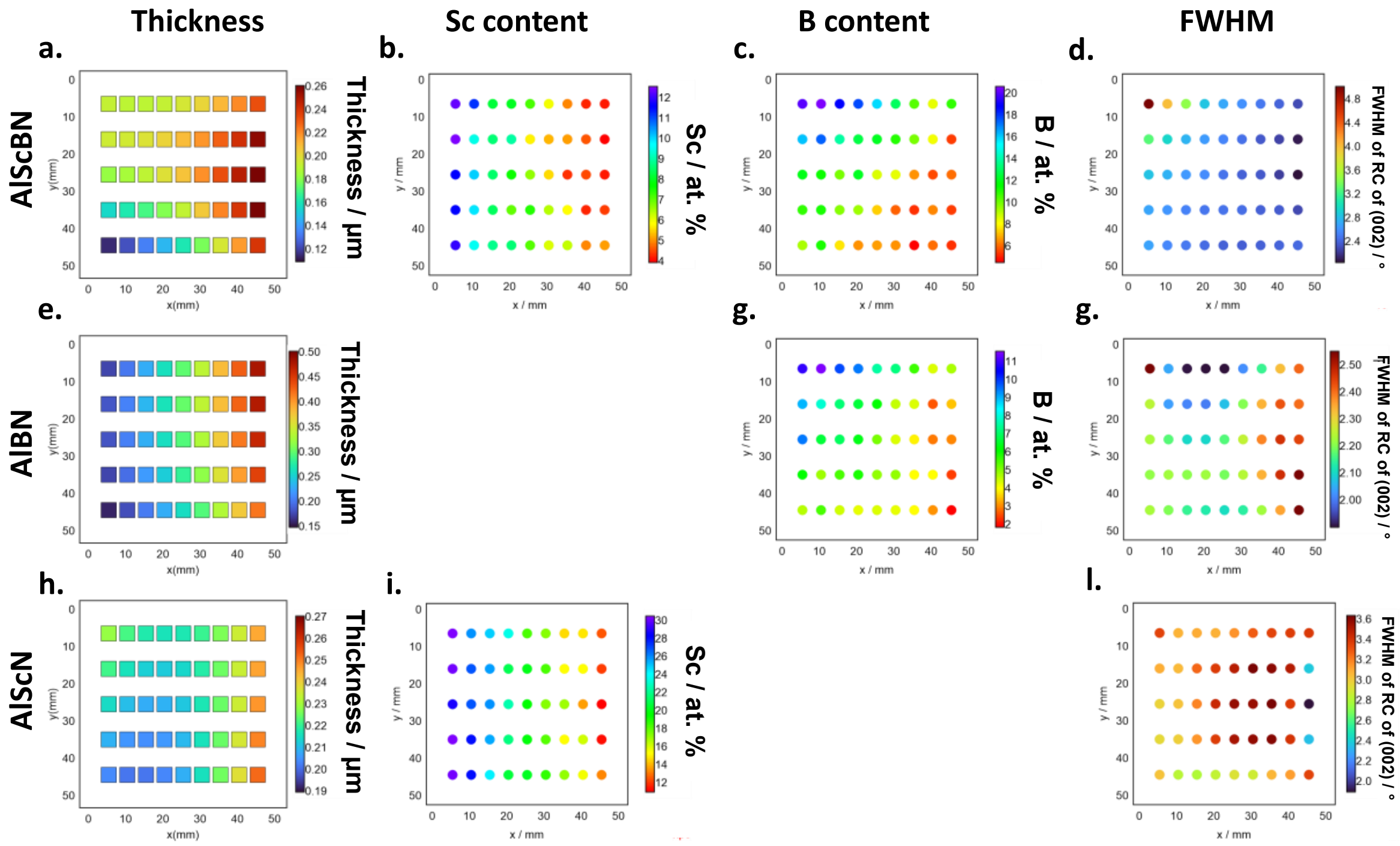

**Figure S3. Thickness and composition analysis:** Representative combinatorial library characterization maps for three material systems. Spatial distribution of film thickness (column 1), Sc content (column 2), B content (column 3), and rocking curve FWHM of the (002) reflection (column 4) across a single representative library for AlScBN (a–d), AlBN (e–g), and AlScN (h–j). Thickness was determined by profilometry, elemental composition by XPS, and FWHM by rocking curve measurements centered on the (002) peak. The systematic compositional and thickness gradients result from stationary substrate positioning during deposition. Higher FWHM values are observed at elevated doping concentrations, approaching the solubility limit. Notably, AlBN films maintain low FWHM values below 2.5° across the full B composition range, consistent with the exceptional crystal quality reported in the main text. Empty panels reflect the binary nature of the respective material systems.

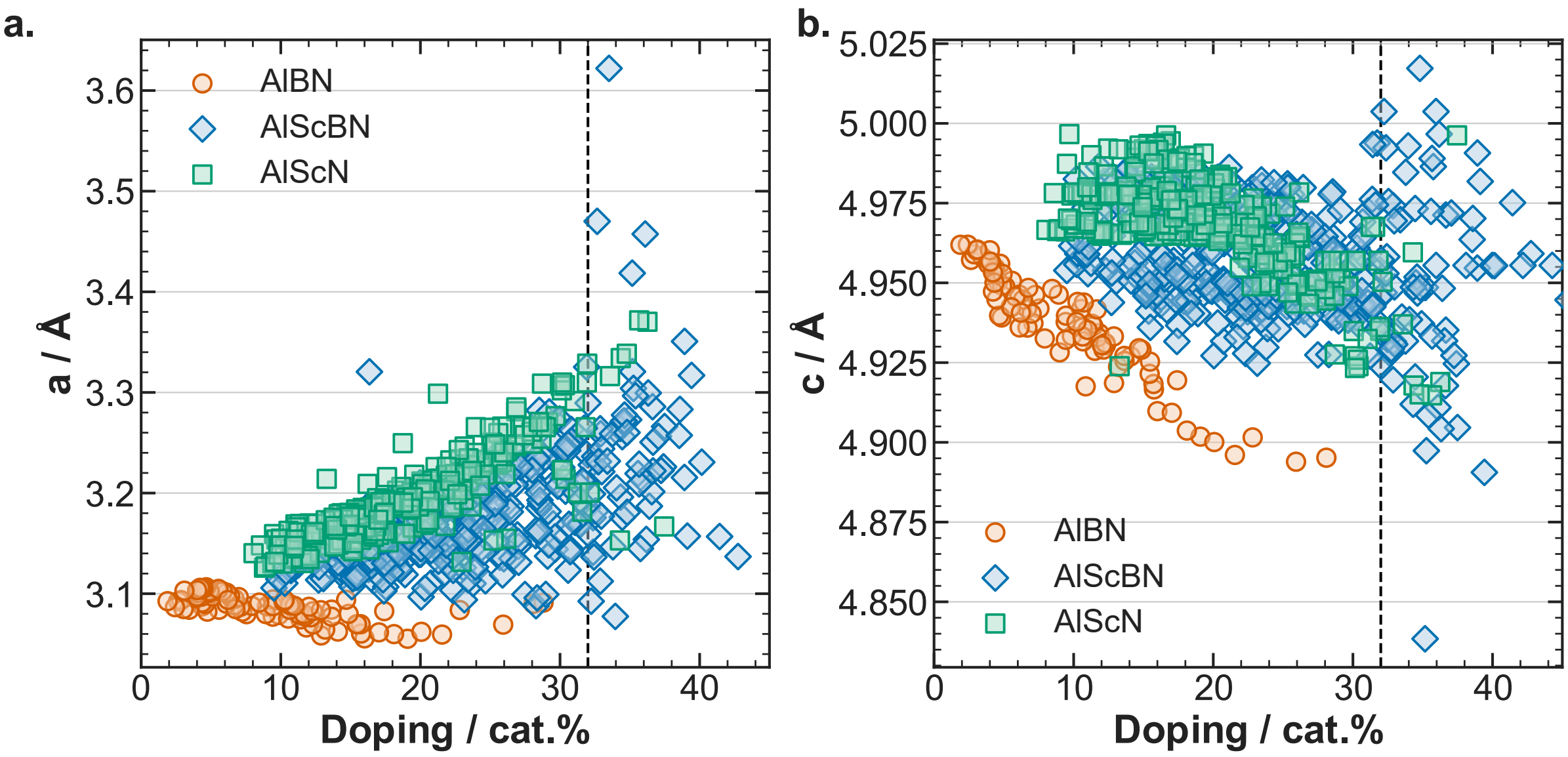

**Figure S4. Lattice parameters of AlScBN, AlBN, and AlScN as a function of total doping content.** (a) In-plane lattice parameter a and (b) out-of-plane lattice parameter c extracted from XRD measurements for all combinatorial library samples. For AlScN and AlScBN, a-parameter increases and c-parameter decreases with increasing doping content, reflecting the well-established lattice distortion driven by the larger ionic radius of Sc relative to aluminum. AlBN shows the opposite trend in both parameters, with both a and c decreasing with doping content, consistent with the smaller ionic radius of B. AlScBN data points fall between the AlScN and AlBN trends across both parameters, confirming the compositional tunability of the lattice. The dashed vertical line at ~33 cat. % indicates the approximate solubility boundary of Sc in binary AlScN for reference; data points beyond this line correspond to compositions approaching or exceeding the single-phase window.

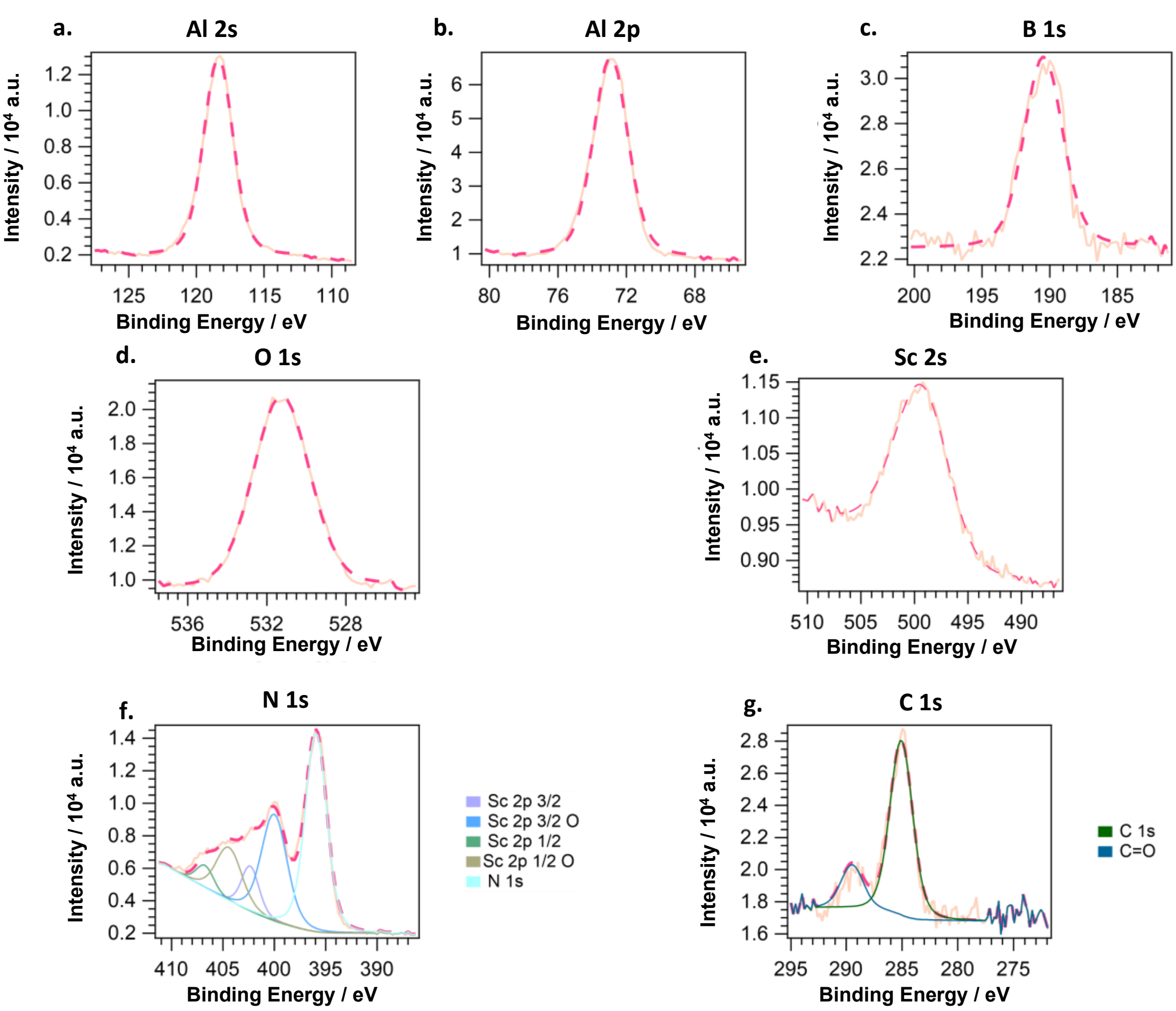

**Figure S5. XPS peak fitting for composition and binding energy analysis.** Representative XPS spectra and peak fitting procedure for AlScBN. Core level spectra illustrating the fitting approach applied consistently across all material systems (AlScBN, AlBN, AlScN): Al 2s (a), Al 2p (b), B 1s (c), O 1s (d), Sc 2s (e), N 1s/Sc 2p region (f), and C 1s (g). Raw data are shown as solid light traces and fit as dashed pink lines. All spectra were referenced to the adventitious carbon C 1s peak at 284.9 eV. The N 1s/Sc 2p region (f) required deconvolution into five components, Sc 2p 3/2, Sc 2p 3/2 oxide, Sc 2p 1/2, Sc 2p 1/2 oxide, and N 1s, due to peak overlap in this binding energy range; a Tougaard background was applied for this region, while a Shirley background was used for all other elements. The C 1s spectrum (g) was fitted with two components corresponding to adventitious carbon and C=O. The O 1s signal (d) is attributed to surface contamination; sputter depth profiling confirmed bulk oxygen contamination below 2 cat. %, with no evidence of bulk oxidation. Cation ratios were calculated from the integrated areas of the Al 2s, Sc 2s and B 1s peaks using their respective relative sensitivity factors.

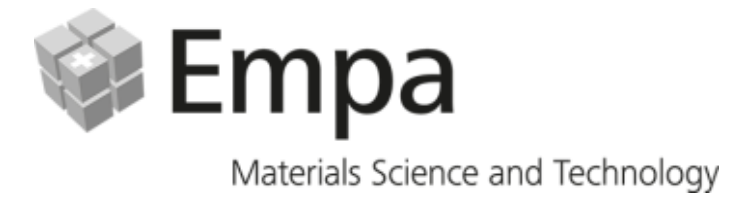

## Ferroelectric characterization of the films

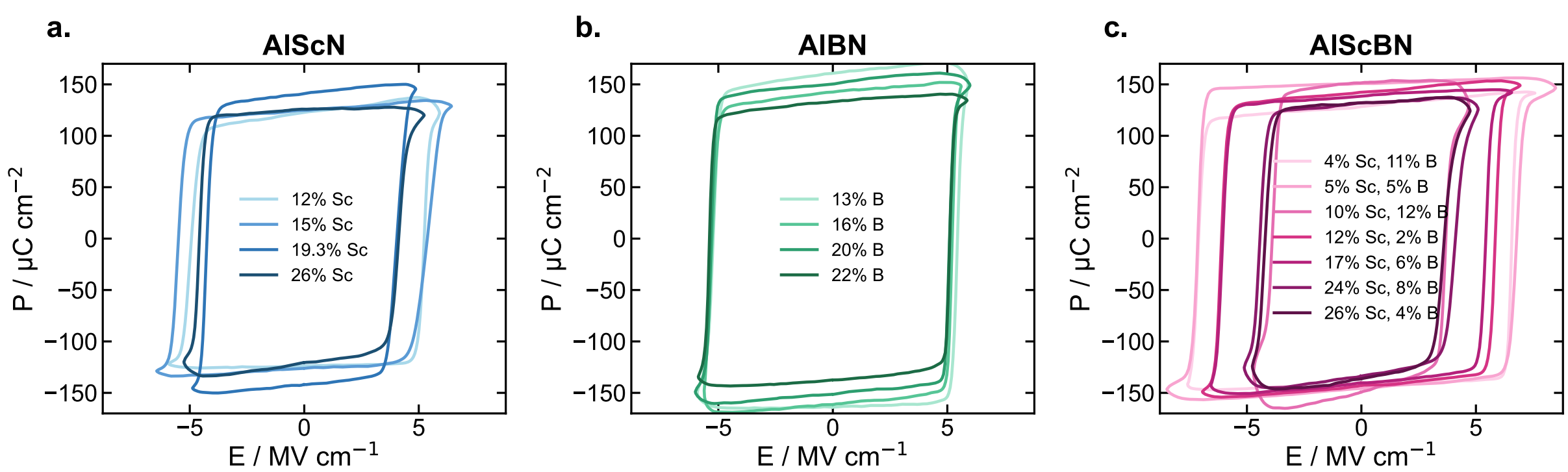

**Figure S6. Representative ferroelectric hysteresis loops for all three material systems.** Polarization-electric field (P-E) loops for AlScN (a), AlBN (b), and AlScBN (c) at selected compositions spanning the ferroelectric window of each material system. All measurements were performed at 3 kHz. Loops are color-coded by Sc content (a, c) or B content (b) as indicated in the legends. The systematic reduction in coercive field with increasing dopant concentration is clearly visible in all three systems, while the remanent polarization remains consistently in the range of 130–150 μC/cm² independent of composition, confirming the trends reported in the main text. The wider applied field range in panel (c) reflects the higher coercive fields observed at low total doping concentrations in the quaternary system.

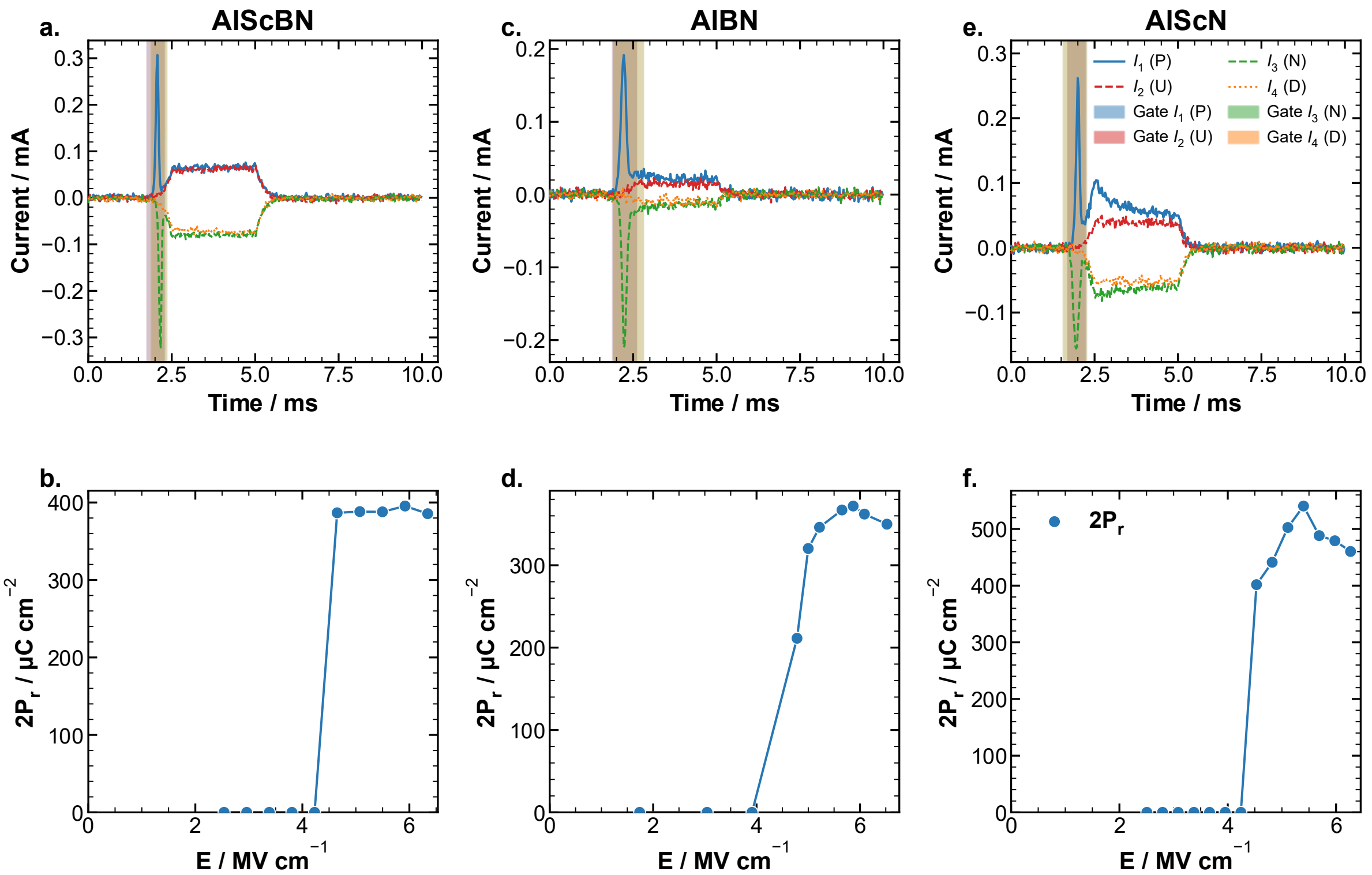

**Figure S7. PUND characterization of representative devices for all three material systems.** Top row: PUND current transients for AlScBN (a), AlBN (b), and AlScN (c), showing clearly resolved switching peaks in $I_1$(P) and $I_3$(N) above the threshold current. The highlighted integration window corresponds to the first switching peak only, isolating the ferroelectric switching current from capacitive charging contributions which dominate at longer timescales. The separation between switching and non-switching pulses confirms genuine ferroelectric switching with quantifiable leakage correction. Bottom row: PUND-corrected remanent polarization $2P_r$ as a function of applied electric field for AlScBN (d), AlBN (e), and AlScN (f). The sharp switching onset followed by clear saturation of $2P_r$ confirms complete polarization reversal in all three material systems. Note that $2P_r$ values reported here are calculated using the nominal pad areas defined by the shadow mask geometry rather than optically corrected contact areas, which may introduce a systematic overestimation of up to 10-15% relative to the corrected values reported in the main text. The switching onset fields are consistent with the coercive field values reported in the main text for each material system.

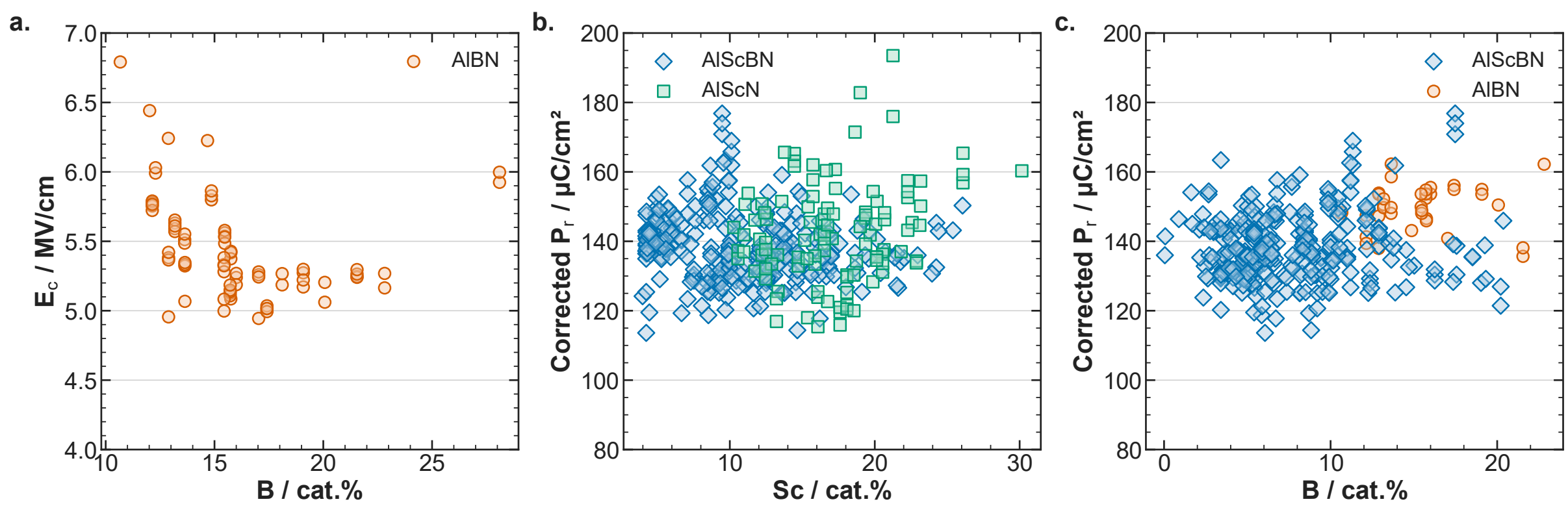

**Figure S8. Coercive field and remanent polarization across the combinatorial library.** (a) Coercive field as a function of B content for AlBN, confirming saturation around 15 cat. % B consistent with previous reports. **(b,c)** PUND-corrected remanent polarization as a function of Sc content **(b)** and B content **(c)** for the indicated material systems. $P_r$ remains consistently in the range of 130–150 µC/cm² across all compositions and material systems, confirming that neither Sc nor B incorporation compromises the ferroelectric polarization within the single-phase compositional window. The dispersion in $P_r$ values is attributed to the pad area correction procedure: while electrode areas were determined by optical microscopy, a single mean area value was applied to all pads within a given die rather than measuring each contact individually. This per-die averaging introduces a residual uncertainty in the normalized polarization values, which manifests as scatter without any systematic dependence on composition.

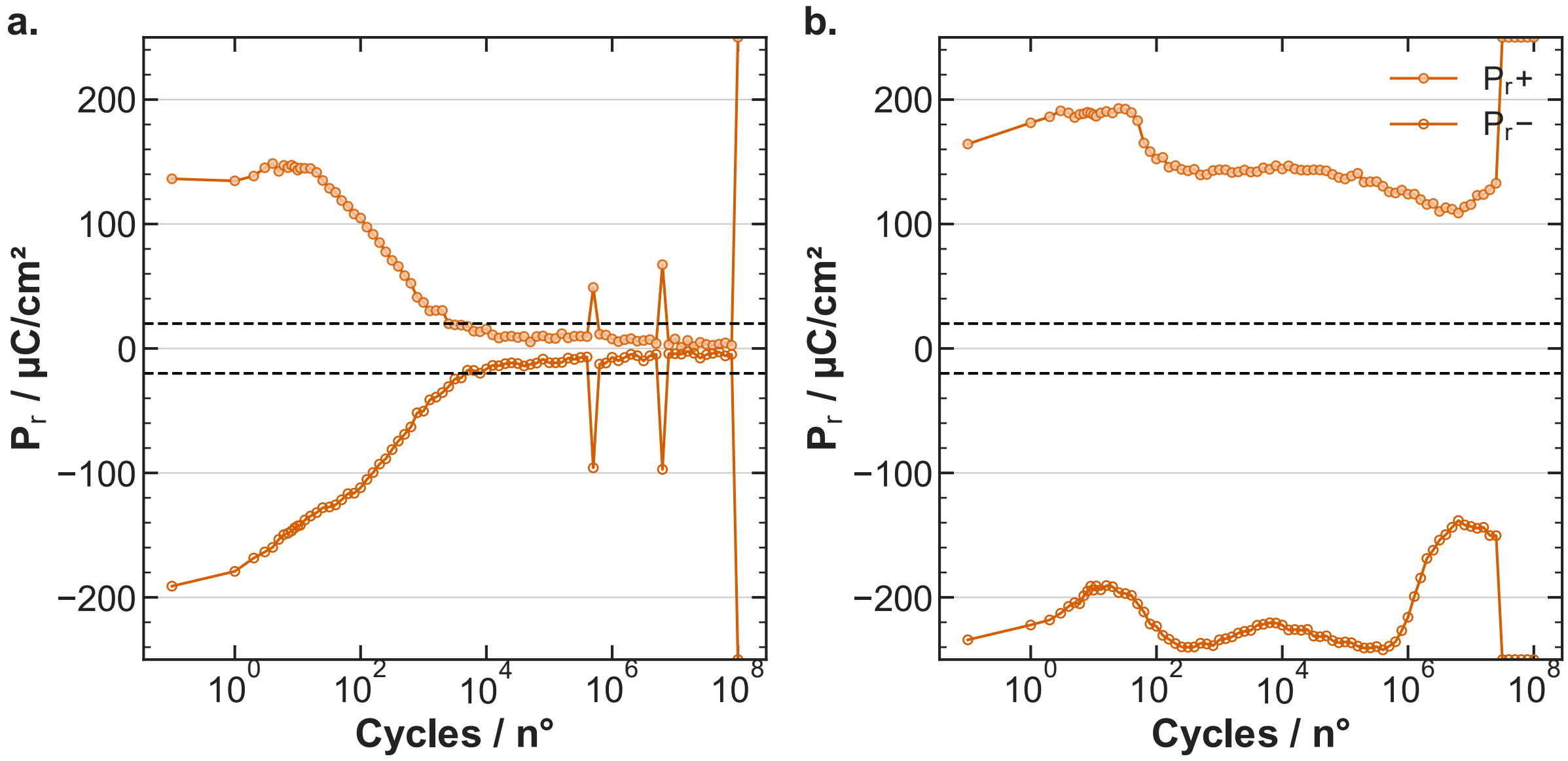

**Figure S9. Ferroelectric cycling fatigue mechanisms in AlBN.** Representative cycling endurance data for two AlBN pads showing polarization $P_r$ as a function of cycle number on a logarithmic scale. (a) Fatigue-type breakdown: $P_r$ decreases gradually until the film depolarizes. (b) Spike-type breakdown: $P_r$ remains stable before a sharp catastrophic event. Dashed lines indicate the ±20 µC/cm² threshold used to define cycling breakdown.

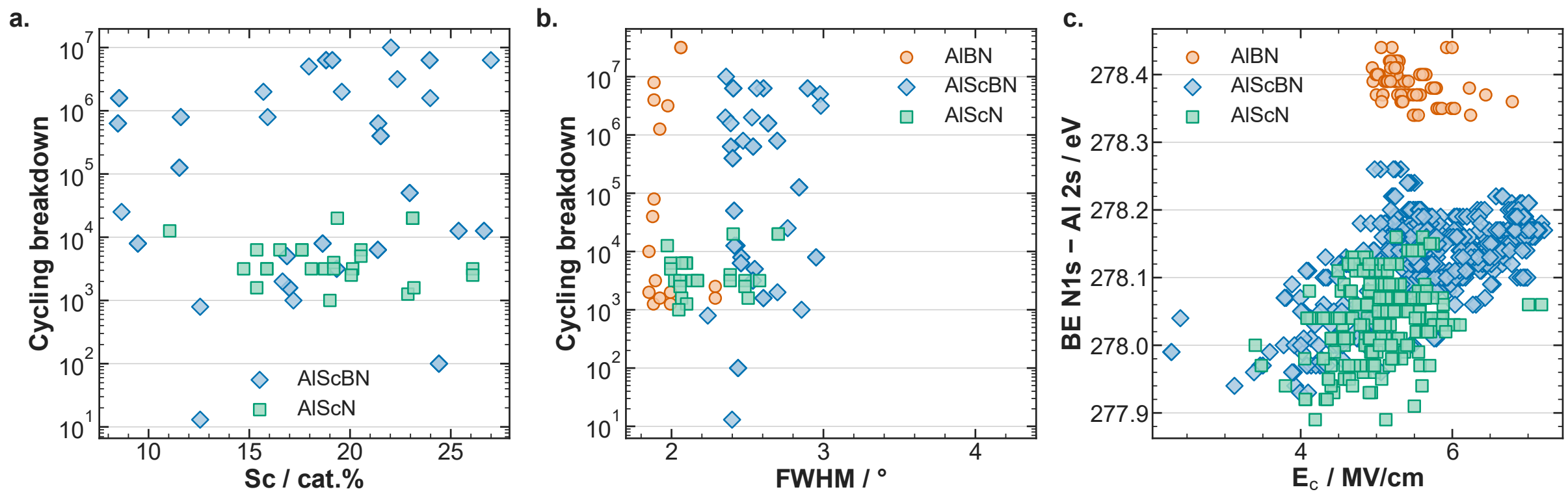

**Figure S10. Structure-property correlations across the combinatorial library.** (a) Cycling breakdown as a function of Sc content for AlScBN and AlScN. No systematic dependence on Sc content is observed within either material system, while AlScBN films show consistently higher endurance than AlScN at equivalent Sc concentrations, confirming that the endurance improvement is driven by B incorporation rather than Sc content. (b) Cycling breakdown as a function of rocking curve FWHM for all three material systems. No clear correlation between crystal quality and cycling endurance is observed, suggesting that endurance is not primarily governed by c-axis texture within the quality range achieved in this study. (c) N 1s – Al 2s binding energy difference as a function of coercive field for all three material systems, demonstrating a direct correlation between bond ionicity and ferroelectric switching field. AlBN forms a distinct cluster at higher binding energy differences, consistent with its fundamentally different switching mechanism discussed in section 2.3, while AlScN and AlScBN overlap strongly, confirming Sc as the primary driver of bond ionicity changes in the quaternary system.